\documentclass[12pt,a4paper]{article}

\usepackage{amsfonts} 
\usepackage{amssymb}
\usepackage[dvips]{graphicx}
\usepackage{epsfig}
\newcommand{\beq}{\begin{equation}}
\newcommand{\eeq}{\end{equation}}

\begin{document}

\hfill SPIN-2003/42

\hfill ITP-UU-03/65

\vspace{30pt}

\begin{center}
{\Large \bf (2+1) gravity for higher genus}\\
\vspace{.3cm}
{\Large\bf in the polygon model}

\vspace{30pt}

{\sl Z. K\'ad\'ar$^{1}$ {\rm and} R. Loll$^{1,2}$}\\

\vspace{24pt}

{\footnotesize
$^1$Institute for Theoretical Physics, Utrecht University,\\
Leuvenlaan 4, NL-3584 CE Utrecht, The Netherlands\\ 
{\tt email:} {\tt z.kadar@phys.uu.nl, r.loll@phys.uu.nl}\\
\vspace{3mm}
$^2$Perimeter Institute for Theoretical Physics,\\
35 King Street North, Waterloo, ON, Canada N2J 2W9\\
{\tt email:} {\tt rloll@perimeterinstitute.ca}
}
\end{center}

\vspace{48pt}

\begin{center}
{\bf Abstract}
\end{center}

We construct explicitly a $(12g-12)$-dimensional space $\cal P$ 
of unconstrained and independent initial data for 
't Hooft's polygon model of (2+1) gravity for vacuum spacetimes 
with compact genus-$g$ spacelike slices, for any $g\geq 2$.
Our method relies on interpreting the boost parameters of 
the gluing data between flat Minkowskian patches as the lengths
of certain geodesic curves of an associated smooth Riemann
surface of the same genus. 
The appearance of an initial big-bang or a final big-crunch
singularity (but never both) is verified for all 
configurations. Points in $\cal P$ correspond to 
spacetimes which admit a one-polygon
tessellation, and we conjecture that $\cal P$ is already the complete 
physical phase space of the polygon model. Our results open the way for
numerical investigations of pure (2+1) gravity.

\vspace{18pt}
  
\section{Introduction}

't Hooft introduced the polygon model of (2+1) dimensional gravity in
order to exclude explicitly the appearance of closed timelike
curves \cite{causality}. Classical spacetimes described by the
model are of the form ${\mathbb R}\times\Sigma$, where the
spatial slices $\Sigma$ may be open or closed two-surfaces 
of any topology. They may also have punctures which
correspond to spinless particles. Pure (2+1) gravity for
spatial slices with torus topology has been studied
exhaustively from many points of view, including its quantization 
(see \cite{carlip} 
and references therein). This is hardly satisfactory, since 
the mathematical structure of the genus-1 case is rather special
and therefore not representative of the general case. 
Although the physical phase space for general genus $g>1$ has
been constructed \cite{witt,mon,hoy}, little is known about the
explicit analytic solutions, or the corresponding quantum theory,
apart from a few results on the genus-2 case \cite{nelson,al}.
Similar difficulties arise when studying coupling to
several point particles (see \cite{cantini,matschull} for
recent progress in this area).

In the present work we will concentrate on the pure gravity case
without particles, and on compact spatial Riemann surfaces
$\Sigma$ of arbitrary genus $g>1$. The cosmological constant is
taken to vanish. Solutions to the classical Einstein equations
have vanishing spacetime curvature everywhere, and are locally
isometric to flat three-dimensional Minkowski space. 
The dynamical variables of the polygon approach are the gluing
homeomorphisms between the flat patches covering spacetime. 
There is a global time parameter, and the associated Cauchy
surfaces of constant time are piecewise flat 
tessellations of $\Sigma$ by polygons.
There are finitely many variables which fix the geometry and
the embedding of the surface, one pair associated with
every polygon edge. They are neither independent nor physical,
since they are subject to a number of constraints and
transform non-trivially under the gauge transformations
of the model. They turn out to obey canonical Poisson brackets if the
Hamiltonian is taken to be the sum of the deficit angles 
around the two-dimensional curvature singularities of the surface 
$\Sigma$ \cite{canquant}.
 
The polygon model is particularly useful in the many instances
where no explicit analytic solution is available
(for example, for $\Sigma$ any Riemann surface of genus $g>1$), since
the classical time evolution is linear and
can easily be simulated on a computer. 
Based on his own simulations, 't Hooft has made conjectures relating the
(non-)appearance of big bang and/or big crunch
singularities to the topology of the spacelike slice, for the case 
of closed $\Sigma$ \cite{evolution}.   
Although the model is derived from the second-order formalism, it can
be viewed as a gauge-fixed version of Waelbroeck's polygon model
\cite{waelmod}, which is a discretized version of the first-order
formalism.
The explicit transformation from the covariant variables of the 
latter to the scalar variables of the 't Hooft model has been given
in \cite{waelthooft}. The quantum theory
arising from the model \cite{canquant} predicts a continuous spectrum
for spacelike and a discrete spectrum for timelike intervals.
Similar results have emerged recently in a loop quantization of
2+1 gravity \cite{rifrro} whose dynamical variables are
closely related \cite{richard}.    

Special cases of the classical polygon model have been studied by
Franzosi and Guadagnini \cite{olaszok}. They solved the torus case 
($g=1$) without particles and described its geometric properties
in detail, and also gave a particular solution of the constraints 
for the higher-genus case. We will extend their work by 
providing explicit solutions for the independent physical initial
data for any genus $g\geq 2$, which we conjecture to be complete.
Our result for the time extension
of the classical solutions coincides with that of 
\cite{olaszok} for the torus case, namely, there is precisely one 
singularity, either initial or final.

The rest of the paper is organized as follows. 
In the next section we review a number of geometric properties of
the 't Hooft polygon model, in order to fix notation and make the
paper reasonably well-contained. 
In Sec.3 we describe how to associate an 
invariant smooth two-surface $S$ with each piecewise flat polygon
tessellation, and then focus on the special case
of a one-polygon tessellation. The time extension of any solution
is proven to be
always a half-line; there is either an initial or a final singularity,
but never both. We describe the action 
of Lorentz symmetry transformations and 
derive an explicit algorithm for solving the constraints 
of the theory, thus arriving at a ($12g-12$)-dimensional space $\cal P$
of freely specifiable, independent initial conditions.  
The solution space $\cal P$ is of the form
${\cal T}_g \times {\mathbb R}_+^{6g-6}$,
where ${\cal T}_g$ is the Teichm\"uller space, the 
space of smooth two-metrics of constant curvature on a genus-$g$ surface modulo 
Diffeo$_0(S)$, the
identity component of the full diffeomorphism group. 
In Sec.4 we generalize part of our construction to
multi-polygon tessellations and spell out our
conjecture that the solution space $\cal P$ obtained from 
one-polygon universes coincides with the complete reduced
phase space of the model. The final
section contains a discussion of our results and an
outlook. Proofs of some of the technical results have been
relegated to four appendices.

\section{Review of the polygon representation} 

Suppose from now on that three-dimensional spacetime is of the form 
$M=\Sigma \times I$ where $\Sigma$ is a
compact orientable surface of genus $g>1$. We will use $g=2$ 
in some of our illustrative examples, 
but our main results hold for any genus $g\geq 2$.
If $M$ is endowed with a locally
flat Minkowski metric, it is a solution to the classical 
vacuum Einstein equations with
zero cosmological constant. One can model $M$ by covering it with
local Minkowski charts and specifying the matching conditions 
\beq 
X'=PX 
\eeq 
between two neighbouring charts $X=(t,x,y)$ and $X'=(t',x',y')$, 
where $P$ is an element of
the Poincar\'e group $ISO(2,1)$ in three dimensions.   
\begin{figure} 
\begin{center}
\includegraphics{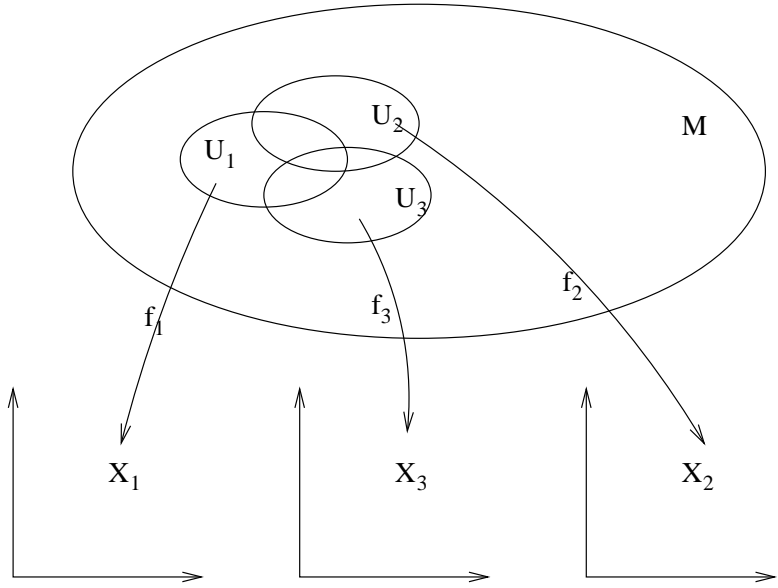}
\caption{\label{walls}{Three overlapping local
neighbourhoods $U_i$ and their associated charts 
on $M$ (diffeomorphisms $f_i:M \to \mathbb{R}^3$);
$f_i \circ f^{-1}_j$ are the Poincar\'e
transformations between neighbouring charts.}}
\end{center}
\end{figure}
Consider now three adjacent patches $U_{1,2,3}$ with coordinate frames
$X_{1,2,3}$ (Fig.\ref{walls}). If the matching conditions are 
\beq 
X_2=P_1 X_1, \quad X_3=P_2 X_2, \quad X_1=P_3 X_3,
\eeq 
in the nonempty
intersection of $U_1 \cap U_2 \cap U_3 \subset M$,
it follows that 
\beq 
P_3 P_2 P_1X_1=X_1. 
\eeq
Writing $P_iX=\Lambda_iX+a_i$ with $\Lambda_i \in SO(2,1)$ and $a_i$ 
a Lorentz vector, we obtain 
\begin{eqnarray} 
\Lambda_3\Lambda_2\Lambda_1 & = & {\bf 1}, \label{lor} \\
                 \Lambda_3\Lambda_2a_1+\Lambda_3a_2+a_3 & = & 0. 
\end{eqnarray}
Every element of the Lorentz group $SO(2,1)$ can be written as
the product of two rotations and a boost, 
\beq 
\Lambda_i=R(\phi_i)B(\xi_i)R(\phi'_i), 
\eeq 
where  
\beq 
B(\xi)=\left( \begin{array}{ccc}
\cosh \xi & \sinh \xi & 0 \\
\sinh \xi & \cosh \xi & 0 \\
0 & 0 & 1 \end{array} \right), \quad 
  R(\phi)=\left( \begin{array}{ccc}
1 & 0 & 0 \\
0 & \cos \phi & -\sin \phi \\
0 & \sin \phi & \cos \phi \end{array} \right). 
\label{rmatrix}
\eeq
Substituting these expressions into (\ref{lor}), one arrives at 
't Hooft's vertex condition
\beq
B(2\eta_3)R(\beta_1)B(2\eta_2)R(\beta_3)B(2\eta_1)R(\beta_2)=1,
\label{vertcond} \eeq
with the identifications 
\begin{eqnarray*} 
\xi_i & = & 2 \eta_i \\ \beta_1 & = & \phi'_3+\phi_2 \\ \beta_2 & = &
\phi'_1+\phi_3 \\ \beta_3 & = & \phi'_2+\phi_1. 
\end{eqnarray*}
The range of the compact angles is $\beta_i \in [-\pi,\pi]$, and 
the factor of 2 in front of the $\eta$'s is a convention which will turn 
out to be useful later. 

\begin{figure}
\begin{center}
\includegraphics[width=8cm,height=7cm]{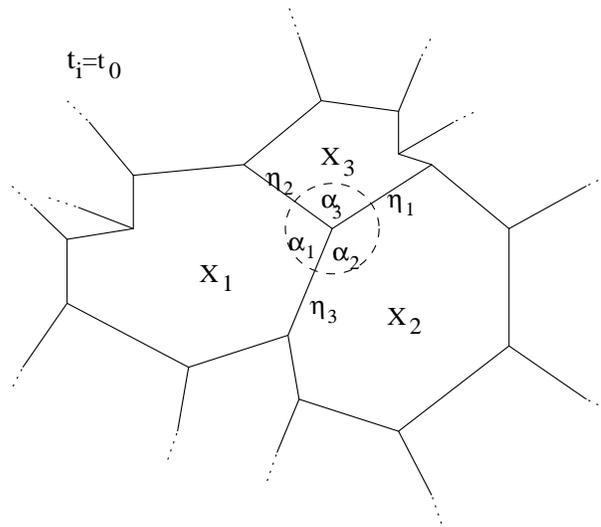}
\caption{\label{timeev}{The geometry of the initial value
surface is characterized by the intersection of the boundaries (\ref{gc})
with the lines $t=const$.}} 
\end{center} 
\end{figure}

The next step is the choice of time slicing. We want
to have a foliation of spacetime with Cauchy surfaces characterized by
a fixed global time coordinate $t$. Fixing  
\beq 
t_1=t_2 \label{gc} 
\eeq
for two adjacent charts $U_1 \cap U_2
\neq \emptyset$ 
defines their common boundary in $M$. For a given time $t^{(0)}$,
setting 
\beq 
t_1=t_2=t_3=...=t^{(0)}
\eeq 
for all the regions (which amounts to 
a partial gauge fixing) defines a piecewise flat Cauchy
surface, where each coordinate
system $X_i$ describes a polygonal region bounded by
the time slices of the hyperplanes $t_i=t_j$ for indices $j$ such that 
$U_i \cap U_j
\neq \emptyset$, as illustrated in Fig.\ref{timeev}. 
We will assume that at most three
polygons meet at each vertex, which presents no loss of generality 
since edges can have zero length at given
instances of time.     
The geometry of the Cauchy surface is completely fixed by the
collection of straight edges of the polygons. They form a
trivalent graph $\Gamma$ with two real parameters associated 
to each of its edges. One is the
boost parameter $\eta$ of the Lorentz transformation in the matching condition
between the two polygons sharing the edge (or of one and the same
polygon in the case of a gluing of two edges from the same polygon).
The other parameter is the length $L$ of the edge. The angles 
$\alpha_i \in [0,2\pi]$ enclosed by pairs of edges incident 
at a vertex are functions of the boost
parameters $\eta_i$ of these edges,
and are related to the rotation angles $\beta_i$ of
eq.(\ref{vertcond}) by
\beq
\alpha_i =\pi -\beta_i.
\eeq  
The explicit dependence is exhibited by writing down the independent 
components of the matrix
equation (\ref{vertcond}) for the variables $\alpha_i$, yielding
\begin{eqnarray}
\label{trirel} 
\cosh (2\eta_k)\!\!\! &=&\!\!\!\cosh (2\eta_i) \cosh (2\eta_j)
+\sinh(2\eta_i)\sinh(2\eta_j)\cos \alpha_k, \cr 
\sinh(2\eta_i)\!\!\!\!\! & : &\!\!\!\!\!\sinh(2\eta_j):\sinh(2\eta_k)
=\sin \alpha_i :\sin
\alpha_j :\sin \alpha_k.  
\end{eqnarray}
In order to avoid misunderstandings, let us
emphasize again that as a result of our choice of time the 
two-dimensional initial value surface $\Sigma$ is flat everywhere except 
at a finite number of vertices $v$ where the sum $\sum_i
\alpha_i$ of the incident angles does not equal $2\pi$.
We have a 2d geometry with conical singularities, where each
singularity contributes with a deficit angle $2\pi -\sum_i
\alpha_i$ to the total curvature, just like in 2d Regge calculus.
By contrast, the three-geometry is {\it flat} everywhere,
including the worldlines of the vertices\footnote{The situation
is different in the presence of
particles, which correspond to real singularities of the three-metric.},
and the flatness condition is precisely eq.(\ref{vertcond}). 

\begin{figure}
\begin{center}
\includegraphics{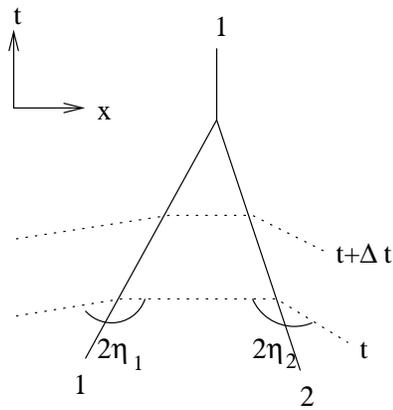}
\caption{\label{trans}{Cut through a piece of spacetime at
constant $y$. Lines of equal time are ``bent''
    at the edges with hyperbolic angles $2\eta_i$. During time
    evolution edges may shrink to zero length and disappear.}}
\end{center}
\end{figure}

\begin{figure}
\begin{center}
\includegraphics[width=6cm,height=3cm]{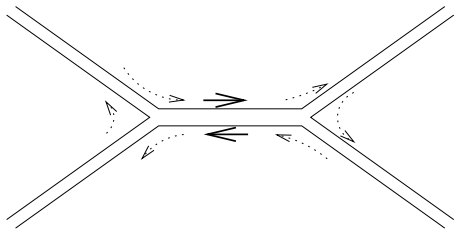}
\caption{\label{ribbon}{Thickening the edges into ribbons, both
sides inherit an orientation from that of the vertices.}}
\end{center}
\end{figure}

Denoting the number of edges in a polygon tessellation by $E$,
the Cauchy problem can be formulated in terms of the $2E$
variables $(L_i,\eta_i)$, $i=1,\dots ,E$. The boost
parameters are constant in time by construction, 
and the evolution of the edges is fixed
by the matching conditions and the gauge condition. Suppose for
simplicity that the origins of $X_1$ and $X_2$ coincide 
so that the matching condition between them is given by
\beq 
R(\phi)B(2\eta)R(\phi')X_2=X_1. \label{sc} 
\eeq
Its first component
\beq 
t_2 \cosh 2\eta +(x_2 \cos \phi'-y_2 \sin\phi')\sinh 2\eta =t_1, 
\eeq 
after imposing the gauge condition $t_1=t_2$ and after elementary
manipulations becomes
\beq 
x_2 \cos \phi'-y_2 \sin\phi'=-t_2 \tanh \eta, 
\eeq 
or, in the other coordinate system
\beq 
x_1 \cos \phi+y_1 \sin\phi=+t_1 \tanh \eta. 
\eeq 
\begin{figure}
\begin{center}
\includegraphics[width=6cm,height=5cm]{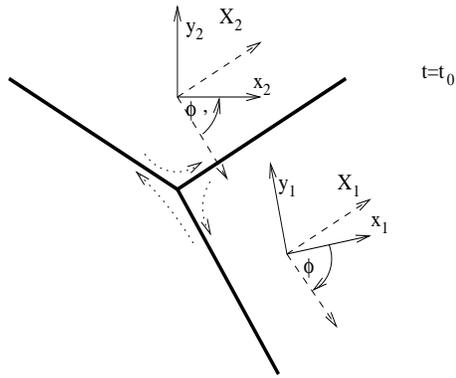}
\caption{\label{matching}{The location of the 
edges (thick lines) is determined by the gauge
and matching conditions.}}   
\end{center}  
\end{figure}
These equations have a straightforward geometric interpretation:
after rotating the coordinate system $X_2$ ($X_1$)
in polygon $2$ ($1$) by an angle $-\phi'$ ($\phi$),
passing across the boundary to the neighbouring 
coordinate system corresponds to a pure boost with 
parameter $2\eta$, as illustrated by Figs.\ref{matching} and 
\ref{trans}. 
The time
evolution of the edges is linear: they move with constant velocity
$\tanh \eta$ perpendicular to themselves and in opposite direction
viewed from the two neighbouring coordinate systems.    
Since $\Sigma$ is orientable, we can give clockwise orientation to
all the vertices as indicated in Fig.\ref{matching}. 
This induces (opposite) orientations on both boundaries of the ribbon when
we thicken out $\Gamma$ into a fat or ribbon graph, as shown
in Fig.\ref{ribbon}.
We can fix the sign ambiguity of the boost
parameters by imposing the convention that $\eta>0$ ($\eta<0$) if
the edges move into (away from) the two adjacent polygons. 
Within this picture, the time evolution is well defined for
an infinitesimal time interval, but the parametrization
can break down whenever an edge shrinks to zero length or a 
concave\footnote{By definition, a convex angle lies between
0 and $\pi$, and a concave one between $\pi$ and $2\pi$.} angle
hits an opposite edge, and part of the variables has to be 
reshuffled. According to the classification in \cite{evolution},
there are nine types of such transitions, during which edges can 
disappear and/or
be newly created. The values of the associated
new boost parameters are unambiguously given by (\ref{vertcond}) and
the requirement that at most one angle can exceed the value of
$\pi$. The violation of this condition would lead to a surface where a
spacetime point is represented more than once, which is not allowed
\cite{gth}.

Since we are interested in studying the time evolution of
2+1 gravity, we must specify a set of initial conditions.
Unfortunately, the variables $(L_i,\eta_i)$, $i=1,\dots,E$,
cannot be specified freely, but are subject to a number of
(initial value) constraints.
There are three constraints per polygon which arise from the
condition that the polygon must be closed. For a polygon with $n$
sides we have
\beq 
\sum_{i=1}^n \alpha_i=(n-2)\pi \label{constr1}
\eeq 
and 
\beq 
\sum_{i=1}^n L_{I(i)} \exp(i\theta_i)=0, \label{comcon} 
\eeq
where $\theta_i=\sum_{j=1}^i (\pi-\alpha_i)$ and $I(i)$ labels the
$i$-th edge starting from a chosen one in counterclockwise 
direction.
We can now count the independent degrees of freedom. 
Starting from $2E$ variables, there are $3F$ constraints,
where $F$ is the number of faces or polygons. 
The number of remaining symmetries is also $3F$, namely, one Lorentz
transformation of the coordinate system at each face. 
However, this action is not free because conjugating 
all Lorentz matrices with the same rotation affects
neither the boost and angle parameters in (\ref{vertcond}), 
nor the lengths; it simply amounts to an overall rotation
of all coordinate systems. We therefore arrive at
\beq 
2E-3F-(3F-1)=-6(F+V-E)+1=-6\chi+1=12g-11 \label{dof} 
\eeq   
for the number of independent degrees of freedom (cf. \cite{olaszok}),
using the formula for the Euler characteristic $\chi=F+V-E=2-2g$ and
$2E=3V$, which comes from the trivalency of $\Gamma$. 
Note that (\ref{dof}) is an odd number.  
As explained in \cite{evolution}, one may choose one of the
length parameters to be ``time'', thus arriving at the usual
$12g-12$ independent parameters defining distinct classical 
universes, the dimension of the reduced phase space of the 
theory \cite{mon,carlip}. Note that in spite of the presence of
conical singularities in the spatial slices, the correct
physical phase space is not the cotangent bundle over 
the moduli space of Riemann surfaces with $p$ punctures 
(which has dimension $12g-12+4p$), but the cotangent space
over the moduli space ${\cal M}_g$ (the space of 
smooth metrics of constant curvature on a genus-$g$ surface modulo
diffeomorphisms\footnote{This moduli space differs from
the Teichm\"uller space ${\cal T}_g$ by an additional quotient
with respect to the discrete mapping class
group action.}), which has dimension $12g-12$. This is so
because the singularities do not correspond to physical
objects in the spacetime, but are merely a consequence of
the gauge choice of the global time parameter.

The difficulty in the polygon representation of 2+1 gravity is
to identify a set of $12g-12$ initial conditions from among the
larger set of $E$ pairs of edge variables $(L_i,\eta_i)$ which
can be specified freely. A major obstacle is the
solution of the constraints (\ref{constr1}), (\ref{comcon}).
This problem has so far not been resolved for $g\geq 2$, although a
particular symmetric solution (with all $L$'s and all $\eta$'s
equal) is known \cite{olaszok}. Note that this problem is
not specific to the polygon representation, but is present
also in other formulations of 2+1 gravity, for example,
in the canonical ``frozen-time'' loop formulation of 2+1
gravity \cite{aetal}. The problem of finding an independent 
set of ``loop variables''
in this formulation was solved in \cite{loll}. 
In the present work, we will present an explicit solution for
the polygon model.

\subsection{Vertex conditions}

As was pointed out in \cite{ruth}, for a certain range of the
parameters involved, (\ref{trirel}) is the
relation between the lengths $2\eta_i$ and the angles $\pi-\alpha_i$ of
a hyperbolic triangle. The triangle inequalities
\beq 
\vert \eta_i \vert+\vert \eta_j \vert \ge \vert \eta_k \vert
\eeq
for all permutations of $(i,j,k)$ edges at a vertex $v$ follow from
(\ref{vertcond}).
One has to be careful when interpreting the boost parameters as
hyperbolic lengths, since one can have negative boost parameters and
concave angles as well. However, by reading eq.(\ref{vertcond}) 
as a 
consecutive action of translations and rotations in hyperbolic space
(the hypersurface $\{(t,x,y)|-t^2+x^2+y^2=-1\}$, with metric 
inherited from 3d Minkowski space), all cases can be 
associated with standard hyperbolic triangles of
positive lengths $\tilde{\eta}_i$ and angles
$0<\tilde{\alpha}_i<\pi$. There are three different cases,
\begin{itemize}
\item{homogeneous vertex:} sgn$\;\eta_i=${\it const}, for which
we set
\begin{eqnarray*} 
\tilde{\eta}_i & = & \vert 2\eta_i\vert , \\
\tilde{\alpha}_i & = & \pi-\alpha_i, 
\end{eqnarray*} 
where $i\in\{ 1,2,3\}$ labels the edges (and opposite angles) 
incident at the given vertex $v$;
\item{mixed vertex:} sgn$\;\eta_3\neq$sgn$\;\eta_2=$sgn$\;\eta_1$, 
where the identification is made according to Fig.\ref{mixedv}, namely,
\begin{eqnarray*} 
\tilde{\eta}_i & = & \vert 2\eta_i\vert ,\\
\tilde{\alpha}_3 & = & \alpha_3-\pi \\
\tilde{\alpha}_i & = & \alpha_i, \quad i=1,2;
\end{eqnarray*}
\begin{figure}
\begin{center}
\includegraphics[width=5cm,height=4cm]{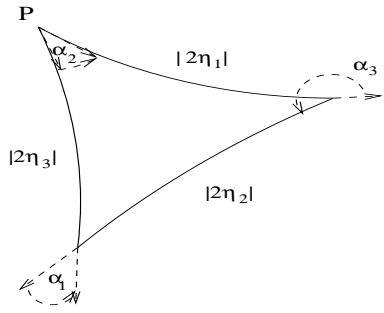}
\caption{\label{mixedv}{Also a ``mixed'' vertex can be associated with
a true hyperbolic triangle with sides $\vert 2\eta_i \vert$, but 
the identification of angles is different from the homogeneous case.
Assuming that $\alpha_3$ is concave, and proceeding clockwise from $P$, 
one reads off
    $\tilde{\alpha}_1=\alpha_1$, $\tilde{\alpha}_2=\alpha_2$ and
    $\tilde{\alpha}_3=\alpha_3-\pi$.}}   
\end{center}
\end{figure} 
\item the degenerate case:
$\vert\eta_i\vert+\vert\eta_j\vert=\vert\eta_k\vert$, 
a limiting case of both of the previous ones, 
characterized by $\sum\tilde{\alpha}_i=\pi$. 
\end{itemize}
(Recall also that in the first case $2\pi<\sum_{i=1}^3\alpha_i<3\pi$ 
and in the
second case $\pi<\sum_{i=1}^3  \alpha_i <2\pi$ \cite{olaszok}.)

\section{One-polygon tessellation and smooth surface} \label{secc}

\begin{figure}
\begin{center}
\includegraphics[width=9.5cm,height=3.5cm]{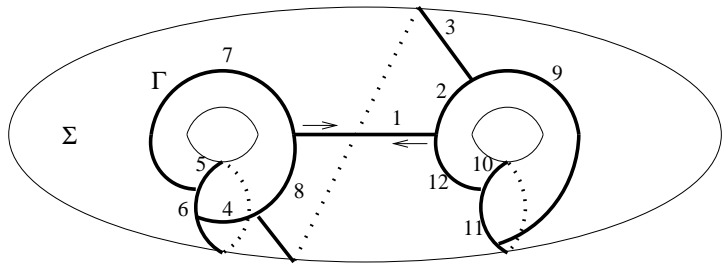}
\caption{\label{tupol}{The spatial universe $\Sigma$ is a piecewise flat 
genus-$2$ surface (thin lines), whose Cauchy data are associated with 
the graph $\Gamma$ (thick lines). The spatial metric on $\Sigma$ is
flat everywhere, but has conical singularities at the vertices
of $\Gamma$, and the edges of $\Gamma$ are straight lines. 
The (thickened-out) graph has two boundary components, 
$(1,2,3,4,5,6,4,8,7,5,6,7)$ and
$(1,8,3,9,10,11,9,2,12,10,11,12)$, and represents a
two-polygon universe.}}    
\end{center} 
\end{figure}

We can decompose the tessellated universe characterized by a graph $\Gamma$ 
into its constituent polygons (disks $D^2_i$ topologically, $i=1,\dots,F$)
by cutting it open along the oriented boundaries of the thickened
ribbon graph, Fig.\ref{ribbon}, which can be expressed by
$\Sigma\backslash\Gamma=\cup_i D_i^2$. An example corresponding 
to a two-polygon universe is given in Fig.\ref{tupol}.
Our construction of independent initial data will involve
an interpretation of the boost parameters $\eta$ 
as geodesic lengths in some hyperbolic space. 
As an intermediate step, this requires the construction
of the graph $\gamma$ dual to $\Gamma$, which is obtained
by choosing a base point $P$ inside each polygon and connecting
the base points of neighbouring polygons pairwise. Since
$\Gamma$ was trivalent, $\gamma$ is a triangle graph on 
$\Sigma$, whose triangles are in one-to-one correspondence
with the vertices of $\Gamma$. The crucial step is to
map $\Sigma$ and its associated curve system $\gamma$ to
a smooth genus-$g$ Riemann surface $S$
of constant curvature $R=-1$ with image graph $\tilde\gamma$.
This is done is such a way that the edges $e_i$ of $\gamma$ 
connecting different base points $\{P_i,\ i=1,\dots,F\}$
or edges connecting a base point with itself
(giving rise to open and closed curves respectively) are mapped 
to geodesic arcs or loops of $\tilde\gamma$ between the image 
points $\tilde P_i$ on $S$.\footnote{In constructing $\tilde\gamma$
explicitly,
one can start by mapping $P_1$ into an arbitrary point $\tilde
P_1$ of $D^2$, and draw one of the arcs starting at $\tilde P_1$ 
in an arbitrary direction (by the homogeneity and isotropy of $D^2$).
The remainder of $\tilde\gamma$ is fixed by the consistent 
$\eta$-assignment and topology of $\Sigma$.}
The (hyperbolic) lengths of these geodesic curves are given by 
(twice the absolute values of) the boost parameters $\eta_i$.
Such surfaces $S$ always exist and are in
one-to-one correspondence with points in the Teichm\"uller 
space ${\cal T}_g$. 

In this section, we will be
dealing with the case of a one-polygon universe,
where a piecewise flat spatial geometry of arbitrary genus
is obtained by making suitable pairwise identifications
of the boundary edges of a single polygon.
We will first prove a number of properties of this
piecewise flat picture, the dual
graph $\gamma$ and the associated smooth surface $S$,
as well as their behaviour under the gauge transformations
of the model. We will then {\it invert} the procedure, by
associating with each surface $S$, together with a certain standard 
geodesic triangulation and
a set of positive real parameters, a unique
one-polygon universe. Since every one-polygon universe
can be obtained
in this way, up to gauge transformations, and since the set
of these universes is closed under time evolution, we obtain
a sector $\cal P$ of dimension $12g-12$ of the
reduced phase space of the model (which has the same dimension).
The generalization to multi-polygon universes is the subject of
Sec.4.

For a one-polygon universe,
the ribbon graph associated with $\Gamma$ has
a single oriented boundary component. The dual graph $\gamma$ consists of 
$6g-3$ closed curves which all begin and end at the same
base point.
An example of a graph $\Gamma$ on $\Sigma$ and the associated
curve system on the corresponding smooth surface $S$ is shown
in Figs.\ref{onepol} and \ref{onepolcur}. 
In the remainder of this section, we will proof various
properties of one-polygon tessellations and their associated
smooth surfaces, by proceeding in a number of steps: \newpage
\begin{figure}
\begin{center}
\includegraphics[width=9.5cm,height=3.5cm]{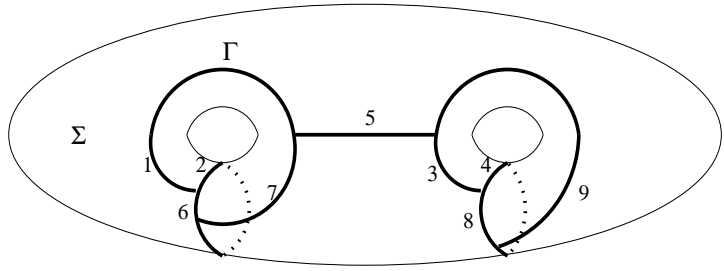}
\caption{\label{onepol}{Example of a graph $\Gamma$
corresponding to a one-polygon universe of genus 2.
}}  
\end{center} 
\end{figure}
\begin{figure}
\begin{center}
\includegraphics[width=9.5cm,height=3.5cm]{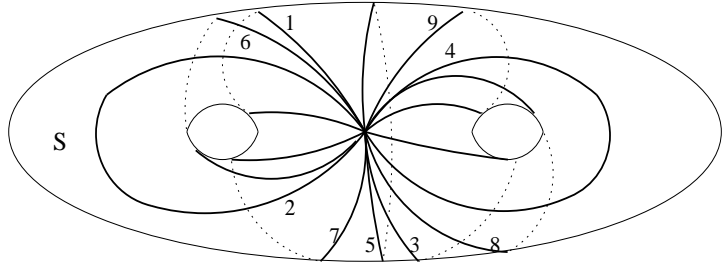}
\caption{\label{onepolcur}{The curve system $\gamma$
corresponding to the one-polygon universe in 
Fig.\ref{onepol} has been mapped to a set of homotopic geodesic loops
on the smooth constant-curvature surface $S$ of the same genus. 
All loops are oriented, and the labels $i=1,\dots,9$ indicate
the outgoing direction from the base point. Note that the picture
is a mirror image (with respect to the paper plane) of the ``correct'' 
one.
}}
\end{center}
\end{figure}
\begin{enumerate}
\item Since a one-polygon universe admits only configurations 
where all $\eta$'s have the same sign, we can identify their (absolute) 
values directly with the lengths of the corresponding geodesic
loops in the triangulation of $S$. For all angles we have 
$\alpha_i=\pi-\tilde{\alpha}_i$, and all are convex. 
These statements will be elucidated further in subsection \ref{op}.
\item The image in $S$ of a triangulation $\gamma$ coming from a 
one-polygon universe can be obtained directly as follows.
After choosing a base point in $S$, take the $2g$ standard
generators $b_j$, $j=1,\dots,2g$,
of the fundamental group $\pi_1(S)$ which have the property that
their complement in $S$ is a geodesic polygon with $4g$ sides 
in the sequence
\beq
b_1b_2b_1^{-1}b_2^{-1}...b_{2g-1}b_{2g}b_{2g-1}^{-1}b_{2g}^{-1}.
\label{funda}
\eeq
The notation $b_k^{-1}$ indicates that the side is to be glued 
with opposite orientation to its partner $b_k$ (with the same $k$)
to obtain $S$ from the geodesic polygon. (An example is given by the 
four loops labeled 1 to 4 in Fig.\ref{onepolcur}.) One then
triangulates the $4g$-sided polygon by drawing geodesic diagonal
arcs until the polygon is triangulated. This is the subject of 
subsection \ref{gp}. 
\item When two out of the resulting $6g-3$ closed loops in $S$
are taken to be {\it smooth} geodesics (ie. without any kinks), 
this uniquely fixes the common base point for all loops to be
the intersection point of the two smooth loops. In
this case only $6g-6$ of the length parameters are independent, and
can be identified with the so-called Zieschang-Vogt-Coldewey 
coordinates of ${\cal T}_g$, as we
discuss in more detail in subsection \ref{zc}.
\item The transitions occurring during the time evolution 
(mentioned in Sec.2 above) correspond to changing the
triangulation by deleting one arc and drawing another one, but
do not alter the surface $S$. This will be explained
in subsection \ref{et}.
 \item The remaining Lorentz symmetry of
the polygon picture corresponds to changing the base point of 
the set of loops on $S$. The length variables $\eta_i$ will transform
in a well-defined way, but the abstract geometry encoded by
$S$ does not change. These properties will be explained 
in subsection \ref{lortra}.
\item Finally, the usefulness of the previous parts of the 
construction will 
become apparent in subsection \ref{compl} where we also solve the constraints 
for the length variables $L_i$, and show that every universe admits 
a one-polygon tessellation
if the Lorentz frame is chosen appropriately.     
\end{enumerate}

\subsection{One-polygon tessellation} \label{op}
If the tessellation consists of a single polygon,
all edges of the dual graph $\gamma$ as well as their
images in $S$ are closed loops which begin and end at
the chosen base point.  
The numbers of triangles and edges are  
\beq 
V=4g-2, \quad E=6g-3, 
\eeq 
as follows immediately from the Euler characteristic and the
trivalency of $\Gamma$. Since the polygon has $n=2E$ sides,
the constraint (\ref{constr1}) reads 
\beq 
\sum_{i=1}^n\alpha_i=(12g-8)\pi, 
\label{anglesum}
\eeq
and for a configuration with sgn$\:\eta_i=${\em const} is equivalent to
\beq 
\sum_{i=1}^n \tilde{\alpha}_i =2\pi.
\label{allsums}
\eeq    
This is true since in the absence of mixed vertices all angles
are convex and we have $\tilde{\alpha}_i=\pi-\alpha_i$ for all angles. 
(Recall that $\tilde{\alpha}_i$ refers to the positive convex angle of
the hyperbolic triangle.) Relation (\ref{allsums}) reflects the fact 
that the base point is a regular point of $S$. 
The proof that sgn$\:\eta_i=${\em const}, $\forall i$, 
is always true for a one-polygon universe is more subtle and
can be found in Appendix A.\footnote{A global reversal of the
signs of the $\eta_i$ corresponds to a reversal of time.}
This has far-reaching consequences.
First, there are only convex angles, so the polygon can never
split. Second,
there is only one transition which can occur, since the other 
eight require either the presence of particles and/or more
polygons. Third, since all boost parameters have the same sign, 
the polygon is either expanding or shrinking at all times. 
In the direction of time corresponding to a shrinking, the
universe always runs into a singularity, because of the lower bound
on the value of the boost parameters (and therefore of the velocities 
of the edges of $\Gamma$), namely, the length of the 
shortest (non-contractible) smooth geodesic
loop on $S$. To summarize, we have found that for all vacuum
universes of the form $\Sigma \times I$ which admit a one-polygon
tessellation, the time interval 
$I$ has to be half of the real axis, $I={\mathbb R}_+$.  

\subsection{Geodesic polygon} \label{gp}
For any triangulation of $S$ by geodesics arising from a triangulation
$\gamma$ dual to a one-polygon tessellation,  
the action of deleting one of its geodesic ``edges'' and instead
inserting the other diagonal of the unique quadrilateral which 
had the deleted edge as its diagonal is called an {\em elementary 
move}. We can use the algorithm given in \cite{mosher} to show that
any triangulation of a given genus-$g$ surface $S$ that arises
in our construction (called an ``ideal triangulation'' in
\cite{mosher}), can be reached
from any other one by a finite sequence of elementary moves. 

\begin{figure}
\begin{center}
\includegraphics[width=7cm,height=7cm]{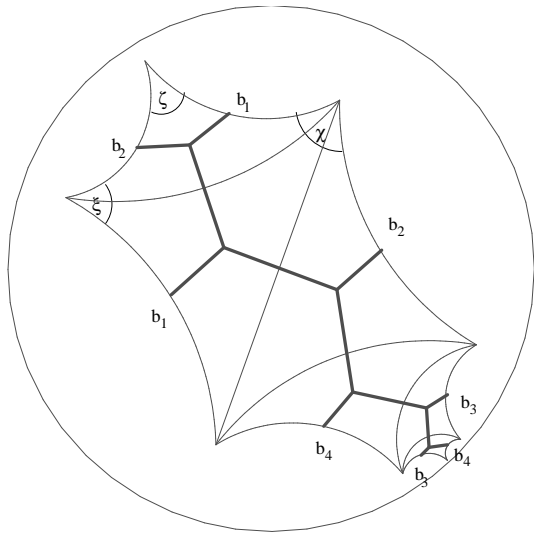}
\caption{\label{ss} {A ``triangulated'' geodesic polygon (thin 
geodesic arcs) representing the smooth surface $S$ on
the unit disk. Edges are to be glued pairwise and with opposite
orientation as indicated by the numbers and in accordance with
relation (\ref{funda}) for the generators of the fundamental
group.
We have also included the original graph $\Gamma$ (thick lines).}}   
\end{center}
\end{figure}

For example, given any graph $\Gamma$, its dual $\gamma$ and a set of
boost parameters $\{\eta_i\}$ for genus two, we can calculate the values of
$\{\eta'_i\}$ corresponding to the curve system in Fig.\ref{onepolcur}. 
The latter has four geodesic loops labeled 1, 2, 3 and 4,
and by cutting the surface along
them one obtains the geodesic polygon with consecutive 
geodesic arcs $b_1b_2b_1^{-1}b_2^{-1}b_3b_4b_3^{-1}b_4^{-1}$,
as explained above.
Fig.\ref{ss} shows the same geodesic polygon drawn on the unit disk 
with standard hyperbolic metric. A similar construction
can be performed for any genus $g$.

\subsection{The ZVC coordinates} \label{zc}

In general the geodesic polygon is described by $6g$ parameters, 
namely, $4g$ angles and $2g$ side lengths. Since all of the angles 
contribute at the same base point $P$, they must sum up to 
$\sum_i\tilde{\alpha}_i=2\pi$. Assume now that $P$ is the intersection 
point of the smooth geodesics labeled 
$1$ and $2$, say, and that these two form part of the triangulation. 
In terms of the geodesic polygon this means
that the two angles at $b_1$ ($\chi$ and $\zeta$ in Fig.\ref{ss}),
as well as the two angles at $b_2$ ($\zeta$ and $\xi$)
should add up to $\pi$. Together with the constraint 
$\sum_i\tilde{\alpha}_i=2\pi$ we therefore have 
three equations for the angles. 
The closure condition for the polygon makes two more
sides and one angle redundant, and we arrive at $6g-6$ degrees
of freedom. It is proven
in \cite{zvs} (see also \cite{buser}) that the so-called 
{\em normal} geodesic polygon
corresponding to the above unique choice of base point
characterizes the surface $S$ uniquely. In other words,
going to the normal polygon amounts to a gauge fixing of
the boost parameters $\eta_i$, which are the lengths of the
arcs of the geodesic polygon, and can be
calculated from the Zieschang-Vogt-Coldewey (ZVC) coordinates 
by using the triangle relations (\ref{trirel}). In Appendix B 
we sketch the algorithm of how
to construct a set of boost parameters corresponding to
an element of ${\cal T}_g$ in practice.

\subsection{The exchange transition} \label{et}

\begin{figure} 
\begin{center} 
\includegraphics{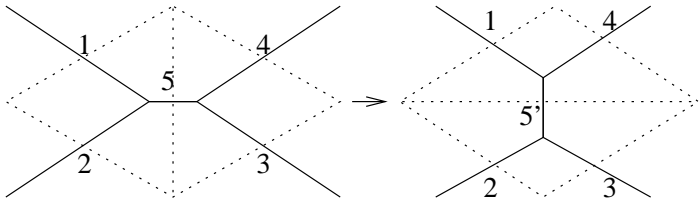}
\caption{ \label{exchange} {During an exchange move, the 
boost parameter $\eta_5$ changes to $\eta_5'$. All new
angles are determined unambiguously. Edges of $\Gamma$ are 
drawn as solid, arcs of the dual triangulation $\gamma$
as dotted lines.}}
\end{center}
\end{figure}

We have seen above how the boost parameters of a 
one-polygon tessellation can be used to
characterize a surface $S$ uniquely. 
However, as we have pointed out already,
within a finite amount of time $\Sigma$ may undergo a transition
which changes both $\Gamma$ and its dual $\gamma$.
In the absence of particles and concave angles only one transition 
can take place, the so-called exchange
transition. As illustrated by Fig.\ref{exchange}, 
the shrinking away of edge 5 of $\Gamma$ (with boost parameter
$\eta_5$) and subsequent ``birth'' of edge 5' (with boost
parameter $\eta_5'$) amounts to a ``flip move'' on the
associated triangulation, namely, the substitution of one
diagonal of a quadrilateral by its dual diagonal,
whose length can be obtained by elementary trigonometry. 
At the level of $S$, this operation corresponds to an
elementary move on the associated triangulation of $S$
as discussed in subsection 3.2 above.
The surface $S$ itself remains unchanged, and is 
therefore invariant under the time evolution. (Note that this
does not imply the absence of time evolution from the original
picture, but only reflects the constancy of the edge momenta or
velocities.) That $S$ is also left invariant by the
residual Lorentz gauge transformations of the polygon model
will be demonstrated in the next subsection.

\subsection{Lorentz transformation}\label{lortra}

An important issue we have not addressed so far is the
role played by the choice of base point in $S$.
As we will show in the following, the action of a Lorentz transformation 
on the boost parameters (that is, a symmetry transformation of
the piecewise flat formulation)
precisely induces a change in the location of the base point. 
Recall the matching condition
\beq 
\Lambda_1X_1+a_1=X_2 
\eeq
between neighbouring coordinate systems $X_1$ and $X_2$
in the polygon picture.
If the edge in question separates two distinct
polygons, the corresponding coordinate systems $X_1$ and $X_2$ can 
be Lorentz-transformed with independent group
elements. In the special case of a one-polygon tessellation we have 
only one coordinate system, and $X_2$ is just an (auxiliary) copy of 
$X_1$. Under the action of a Lorentz transformation $\Lambda$ 
we have $X_j\to \tilde X_j=\Lambda X_j$, $j=1,2$, and the matching 
condition gets modified to
\beq 
\tilde\Lambda_1 \tilde X_1+ \tilde a_1=\tilde X_2 
\label{lortilde}
\eeq 
with $\tilde\Lambda_1=\Lambda\Lambda_1\Lambda^{-1}$ and 
$\tilde a_1=\Lambda a_1$.

To see the effect of this transformation on the triangulation of $S$, 
it is convenient to use a different representation
of $S$. Any Riemann surface $S$ with genus $g>1$ can
be represented as a quotient $H/G'$, where
$H$ is the hyperboloid $\{(t,x,y)| -t^2+x^2+y^2=-1\}$ with $ds^2$
inherited from the three-dimensional Minkowski space and 
the subgroup $G'\subset SO(2,1)$ is
isomorphic to the fundamental group $\pi_1(S)$. 
For simplicity we will use the isomorphic quotient $D^2/G$ instead, 
where $D^2$ is the open unit disk with line element
\beq 
ds^2=\frac{4\vert dz\vert^2}{(1-\vert z\vert^2)^2} 
\eeq
and $G\subset$ {\it PSU}$(1,1)$ 
({\it PSU}$(1,1)$ is isomorphic to $SO(2,1)$). 
The group action on points $z$ of the unit disk is given by
\beq 
z \mapsto gz=\frac{g_{11}z+g_{12}}{g_{21}z+g_{22}}. \label{su11} 
\eeq 
We use {\it PSU}$(1,1)=SU(1,1)/\mathbb{Z}_2$, because any element
$g \in SU(1,1)$ has
the same action on the coordinate $z$ as $-g$. 
Identifying boost and rotation matrices,
\beq 
b(\xi):=\left( 
\begin{array}{cc} \cosh\frac{\xi}{2} & \sinh\frac{\xi}{2} \\
                  \sinh\frac{\xi}{2} & \cosh\frac{\xi}{2}
\end{array} \right), \quad
 r(\phi)\:=\left(
\begin{array}{cc} \exp(i\frac{\phi}{2}) & 0 \\
                         0              & \exp(-i\frac{\phi}{2})
\end{array} \right), 
\eeq
any {\it PSU}(1,1) matrix can be written as 
\beq
g=r(\phi)b(\xi)r(\phi'), \label{dec} 
\eeq
and the isomorphism with SO(2,1) is given by $B(\xi) \leftrightarrow
b(\xi)$ and $R(\phi) \leftrightarrow r(\phi)$, cf. (\ref{rmatrix}). 
Consider now a universal cover $f:D^2 \to S$ of the surface
$S$, and let $z_0$ be an inverse image of the
base point $P\in S$. Consider the $2g$ lifts of the geodesic loops 
corresponding to the $2g$ standard generators $b_i$ of $\pi_1(S)$. 
There is a unique element $g_i$ in
$G$ which maps $z_0$ to the other end point of the lift of
$b_i$, which is a geodesic arc in $D^2$. 
When decomposing the group element $g_i$ according to
eq.(\ref{dec}), the boost parameters are identified with 
the parameters $\xi_i$ (the lengths of the arcs connecting
$z_0$ and $g_iz_0$) according to $2\eta_i=\xi_i$.
The elements of the Teichm\"uller space ${\cal T}_g$ are in 
one-to-one correspondence with the
conjugacy classes of discrete subgroups of $PSU(1,1)$, 
that is, $G$ and $hGh^{-1}$ with
$h \in${\it PSU}(1,1) determine the same surface $S$. 
If the base point for the action of $G$ was
$f(z_0) \in S$, it is $f(hz_0)$ for $hGh^{-1}$, where the 
generators of $hGh^{-1}$ are simply obtained by conjugating the
standard generators with $h$. Like the action of the corresponding
$\Lambda\in SO(2,1)$ by conjugation (defined below (\ref{lortilde})),
the action of $h\in${\it PSU}(1,1) has a non-trivial (one-dimensional) 
isotropy group.
This completes the argument that a Lorentz transformation on 
the coordinate system associated with the polygon translates
into a change of the base point of the triangulation of
$S$, while leaving the surface $S$ itself unchanged.

\subsection{The complex constraint}\label{compl}

While we have found an abstract geometric re-interpretation for the 
boost parameters of the polygon representation, which has enabled
us to identify the independent physical and the 
redundant gauge degrees of freedom, nothing has been said so far
about the other half of the canonical variables, the edge length
variables $\{ L_i\}$. We will show in the following that
the complex constraint (\ref{comcon}) admits a
solution for any triangulation of any surface $S \in {\cal T}_g$, 
provided that the base point $P\in S$ is chosen carefully. 
Furthermore, we will show that from a particular solution 
(of eq.(\ref{hsz}) for all relevant triplets {$z_i$})
one can construct a $(6g-6)$-parameter family of solutions, 
thus spanning an entire sector $\cal P$ of the full phase space 
where ${\cal P}= \mathbb{R}_+^{6g-6}\times {\cal T}_g
\stackrel{\rm diff}{=}
 \mathbb{R}^{12g-12}$. 

For the case at hand, we can rewrite the complex constraint
(\ref{comcon}) as
\beq 
\sum_{I=1}^{6g-3} L_{I}z_I=0 \label{comcon2} 
\eeq
with $z_I=\exp(i\theta_{i})+\exp(i\theta_{j})$,  
since each label I will appear exactly twice, namely, at positions
$i$ and $j$ when counting the edges of the polygon in 
counterclockwise direction. 
The angles $\theta_i$
are expressible as sums of angles $\alpha_i$ of the polygon, but 
there is a more straightforward way of writing them. 
\begin{figure}
\begin{center}
\includegraphics{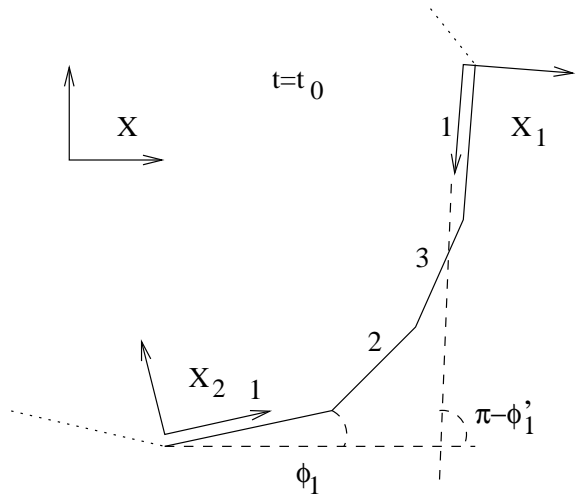}
\caption{\label{teta} {Starting from the edge
labeled $1$ and proceeding in counterclockwise direction 
one has $I(1)=1,\:I(2)=2,\:I(3)=3,\:I(4)=1$, $\dots$ and therefore
$z_1=\exp(i\theta_{1})+\exp(i\theta_{4})$. Instead of adding the
outer angles in terms of $\alpha_i$ to obtain $\theta_i$, we can
determine them directly from the angle parameters appearing in the
Lorentz transformation part of the matching conditions
corresponding to the edge $1$.}}   
\end{center}
\end{figure} 
If the matching condition was $X_2=\Lambda_1X_1+a_1$ with 
\beq 
\Lambda_1=R(\phi_1)B(2\eta_1)R(\phi'_1), 
\label{matchagain}
\eeq 
one can rewrite it as $R(-\phi_1)X_2=B(2\eta_1)R(\phi_1')X_1+a'_1$, 
and Fig.\ref{teta} then shows that 
\beq 
\theta_1=\phi_1, \quad \theta_4=\pi-\phi'_1. 
\eeq
In order to derive this relation, one has to take into account
that (after an appropriate translation) $X_2$ has to be rotated
by an angle $-\phi_1$ and $X_1$ by an angle $\phi'_1$ to align
their spatial axes with those of $X$, and that the two occurrences
of an edge always have opposite orientation.
After rotating the spatial axes of $X_1$ and $X_2$ to those of $X$, 
the matching condition between the new
coordinate systems is a pure boost $B(2\eta_1)$.
The coefficient $z_1$ of the corresponding edge 
is therefore given by
\beq 
z_1=\exp(i\phi_1)+\exp(i(\pi-\phi'_1)). 
\label{zcoeff} 
\eeq  
The constraint (\ref{comcon2}) is a complex linear equation, and 
has a solution if and
only if the complex coefficients $z_i$ -- thought of as vectors based at
the origin of the complex plane -- are not contained in a half 
plane\footnote{Note that the degenerate case where all $z_I$'s are
collinear cannot occur.}. If
they are, there is no non-trivial linear combination with
positive coefficients $L_i$ which vanishes. 
In Appendix C we prove the non-trivial fact that after an
appropriate conjugation of the generators $g_i$ of $G$ which correspond 
to the
loops of the triangulation, the coefficients $z_i$ will be transformed
into a generic position, not contained in any half plane. 

\begin{figure}
\begin{center}
\includegraphics{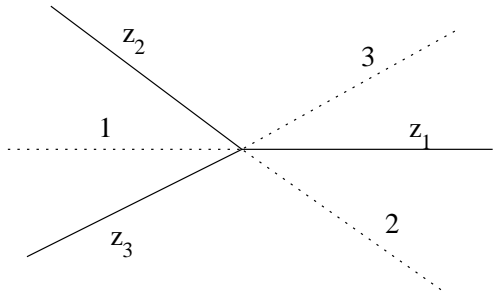}
\caption{\label{sol} {For any point $z_i$ ($i>3$) in one of the convex
sections of the complex plane bounded by two half lines 
$j,k \in\{1,2,3\}$, the triangle with corner points $(z_i,z_j,z_k)$ 
contains the origin. In other words, the corresponding eq.(\ref{hsz}) 
has a unique solution with $\lambda_j,\lambda_k,\lambda_j+
\lambda_k \in ]0,1[$.}}   
\end{center}
\end{figure}

Suppose now that this has
been achieved, ie. there are three points $z_1$, $z_2$, $z_3$ in
the desired generic position. We can divide the complex plane as 
depicted in Fig.\ref{sol}, and $z_i$, $ i>3$, is some other point 
lying in the
convex section of the plane bounded by lines $1$ and $3$, say. 
Then the equation
\beq 
\lambda_1 z_1+\lambda_2 z_3+(1-\lambda_1-\lambda_2)z_i=0
\label{hsz} 
\eeq
admits a unique solution with $\lambda_1,\lambda_2, \lambda_1+\lambda_2 
\in ]0,1[$.
A similar statement holds for any point $z_i$ contained in one 
of the other two convex sections of the plane. There are $6g-5$
independent triangles ($(1,2,3)$, and every other index 
$i\in [4,6g-3]$ matched with two of $(1,2,3)$ according to the location
of $z_i$ as explained above). Adding up the resulting $6g-5$
equations of the form (\ref{hsz}), each one multiplied by an arbitrary
number $\rho_i>0$, we get a solution to the constraint 
(\ref{comcon2}). 
Each index is represented, and each $z_i$ appears with a positive 
coefficient $L_i$, namely, a positive
linear combination of the $\rho_j$. All $\rho_i$'s are
independent, but we can fix $\rho_1$ to be $1$, which fixes
the global time parameter or, equivalently, an overall length scale.
Thus we have completed the explicit
construction of a $(12g-12)$-parameter set $\cal P$ of independent and 
unconstrained initial conditions for the polygon model, each
corresponding to a one-polygon universe. We have obtained
this space in the explicit form 
${\cal P}={\mathbb R}_+^{6g-6}\times {\cal T}_g$. 
We conjecture in Sec.4 below 
that $\cal P$ is identical with the full phase space
of the model, and not just an open subset of it.

\section{Multi-polygon tessellation and smooth surface}\label{multip}

What we have described up to now is the sector of the
theory corresponding to a single polygon. For this case, we
have identified a complete set of initial data (the phase
space $\cal P$), and shown that it is mapped into itself
under time evolution. However,  
as we have mentioned in the introduction,
a generic universe in the 't Hooft representation is a whole
collection of flat polygons glued together at their boundaries. 
We will in the following prove a number of results about
multi-polygon configurations, and discuss their relation to the
one-polygon sector of the theory. 
Because of technical problems to do with the complicated
action of the Lorentz gauge transformations on these 
configurations, we have so far been unable to establish 
explicit solutions to the analogues of the constraint
equation (\ref{comcon2}) and to construct a complete set of
initial data. These questions may ultimately turn out to be
irrelevant if a conjecture of ours is correct, namely, that
any multi-polygon universe is physically equivalent to one
in the one-polygon sector described in the previous section. 

A universe consisting of $F$ polygons has a graph $\Gamma$
with $6g+3(F-2)$ edges. Since every edge comes with a
canonical variable pair $(\eta_i,L_i)$, it is clear that
$3F$ of these pairs must correspond to unphysical or
redundant information. For $F>1$, one would expect that
at the level of the $\eta$'s alone three more boosts can
be gauge-fixed for every additional polygon in the tessellation.
We will illustrate below by a specific example how a multi-polygon
universe can be effectively reduced to a universe with fewer
polygons by a suitable gauge-fixing.
   
We will begin by explaining how 
our method of associating a unique set of Teichm\"uller parameters
to every universe generalizes to the case of more than one
polygon. At the level of the piecewise flat physical Cauchy surface 
$\Sigma$, the dynamics of a multi-polygon
tessellation involves also other types of transitions 
\cite{evolution} which in general will change $F$, and which 
we will describe next.
As a new feature of the multi-polygon case, we will see that the
boost parameters can now have different signs.
Finally, we will discuss gauge-fixing and formulate our
conjecture.

\subsection{Multi-polygon tessellation}\label{mpt}

We have already explained in the introductory part of Sec.3
how to associate a unique smooth uniformized surface with a geodesic
triangulation $\tilde\gamma$ to every pair $(\Sigma,\Gamma)$.
Like the dual graph $\gamma$, $\tilde\gamma$ still has $F$
base points and $6g+3(F-2)$ geodesic ``edges". 
Since the set of $\eta$'s carries a redundancy, one cannot
read off directly the Teichm\"uller parameters which
characterize the hyperbolic surface $S$ uniquely. 
However, there is a straightforward
procedure to effectively reduce the graph $\tilde\gamma$ to 
that of a normal geodesic polygon (cf. Sec.3.3 above),
with a minimal number $6g-3$ of edges.
It involves the ``gauging away" of a number of geodesic arcs
which link different base points $\tilde P_i$, so they can
all be thought of as one and the same base point corresponding
to a one-polygon configuration. 
Details of the procedure can be found in Appendix D.

\subsection{Transitions for multi-polygons}\label{tfmp}

For the case $F>1$, four more transitions can occur during
the course of the time evolution, 
in addition to the ``exchange move'' already discussed in
Sec.3.4. The new transitions are qualitatively different 
since individual edges of $\Gamma$ may collapse to zero
length, or new edges may be generated when a vertex at a
concave angle hits the opposite side of a polygon.
Those of the transitions which change the number
$F$ of polygons (and therefore of edges) are
awkward to deal with, since they correspond to ``jumps''
in the number of dynamical variables {\it before} taking
constraints and gauge symmetries into account.
The evolution of the {\it physical} phase
space variables is of course perfectly well-defined and unique
during these transitions as was shown in \cite{evolution,gth}, 
but this is not the form in which
the theory is given in the first place. 

Let us enumerate the new transitions one by one and study what 
effect they have on the dual triangulation $\tilde\gamma$. 
The transitions will also play a role when considering finite
gauge transformations, cf. subsection 4.3 below.

\begin{figure} 
\begin{center}
\includegraphics{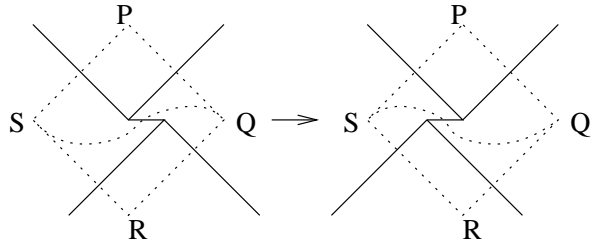} 
\caption{\label{vg} {Vertex grazing.}}
\end{center}
\end{figure}

\begin{figure} 
\begin{center}
\includegraphics{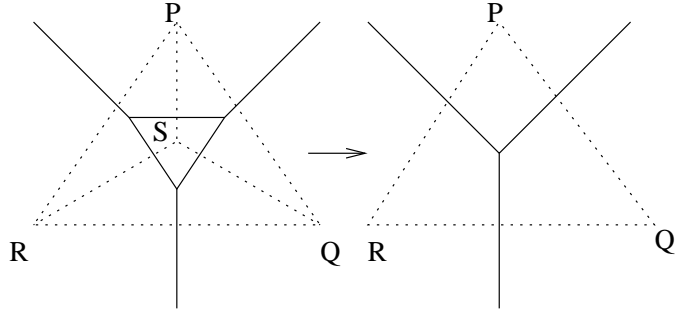}
\caption{\label{td} {Triangle disappearance.}}
\end{center}
\end{figure}

\begin{figure} 
\begin{center}
\includegraphics{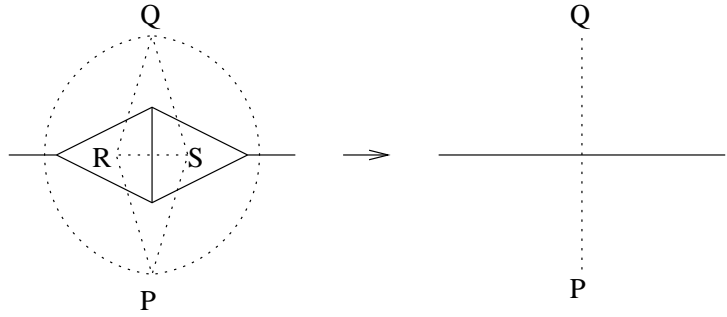}
\caption{\label{dtd}{Double triangle disappearance.}}
\end{center}
\end{figure}

\begin{figure} 
\begin{center}
\includegraphics{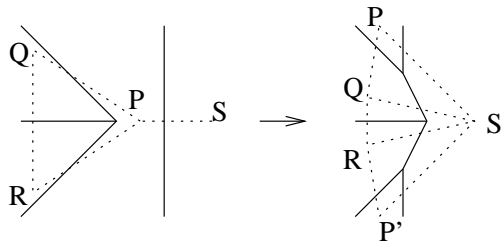}
\caption{\label{vh} {Vertex hit.}}
\end{center}
\end{figure}

\begin{enumerate}
\item{vertex grazing:} $F \to F,\;\eta \to -\eta$

As is clear from Fig.\ref{vg}, this transition does not change
the triangulation $\gamma$, in agreement with the fact that the
order of boundary edges around each polygon remains the same. 
The absolute value of the boost parameter of the shrunk 
edge does not change, but its sign does.

\item{triangle disappearance:} $F \to F-1$

The time reversal of this transition gives a refinement of the 
triangulation $\gamma$. Fig.\ref{td} shows that the 
corresponding base point $S$ disappears. The three dual edges
emanating from $S$ are therefore ``deleted'', as well as
the corresponding edges of the polygon(s).  

\item{double triangle disappearance:} $F \to F-2$

Like in the previous transition, the time reversal 
of this one is also a refinement, as illustrated in Fig.\ref{dtd}. 
Note that the two curves connecting $P$ and $Q$ lie in the same
homotopy class in $S$
so their corresponding (absolute values of) boost
parameters must coincide even though they correspond to
different edges of $\Gamma$. 

\item{vertex hit:} $F \to F+1$

This transition corresponds directly to a refinement of the 
triangulation $\tilde{\gamma}$. Fig.\ref{vh}
shows a somewhat misleading picture, since the points $P$ and $P'$ 
are actually the same, but correspond to different polygons. In other
words, $PS$ and $P'S$ are different curves in $\gamma$ but
coincide in $\tilde{\gamma}$ on $S$.\footnote{We need to
keep both points nevertheless. For example, an exchange of the edge
corresponding to $PS$ in $\Sigma$ will result in erasing $PS$ and
drawing instead the other diagonal of the quadrilateral whose diagonal 
was $PS$, while leaving $P'S$ intact.} The three new edges
are the arcs $QS$, $RS$ and $P'S$, a duplicate of $PS$. 

\end{enumerate}  

\subsection{Reducing multi-polygons by gauge-fixing}\label{reduce}

\begin{figure} 
\begin{center}
\includegraphics{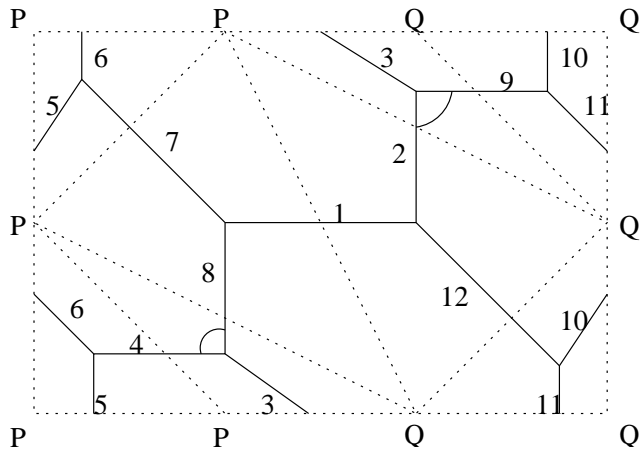}
\caption{{\label{tupoltri} Alternative schematic representation 
of the $F=2$ universe corresponding to Fig.\ref{tupol}, showing
both the graph $\Gamma$ (solid lines) and its dual
$\gamma$ (dotted lines). 
The boundary edges of the rectangle (corresponding to smooth 
geodesic arcs) must be glued pairwise with opposite
orientation, such that the open edges of $\Gamma$ 
are glued pairwise according to matching numbers. 
In the corresponding graph $\tilde\gamma$ on $S$,
moving $P$ to $Q$ along the dual edge $3$ results in the degeneration 
of triangles $348$ and $239$. Simultaneously, 
the two polygon angles indicated on the figure as well as the sums of
the other two angles at both vertices  
become $\pi$, edge 3 becomes redundant, the two vertices do not carry deficit
angle any more and 48 and 29 can be
thought of as the same edge. The configuration becomes effectively
that of a one-polygon universe.
}}
\end{center}
\end{figure}

\begin{figure} 
\begin{center}
\includegraphics[width=9cm,height=4.5cm]{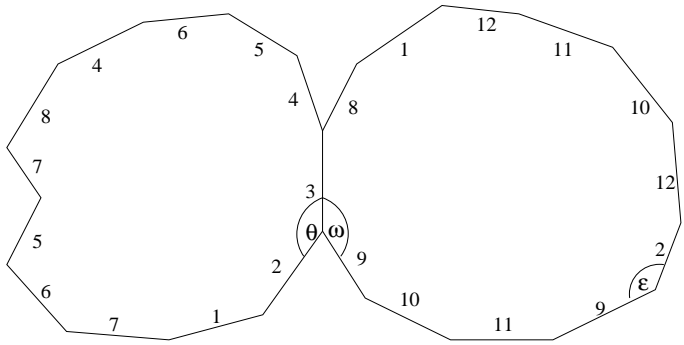}
\caption{\label{tupolfla} {The polygons corresponding to
the two-polygon universe of Figs.\ref{tupol} and
\ref{tupoltri}. If one Lorentz-transforms the two coordinate frames
such that $\eta_3$ becomes zero, the angles $\theta+\omega$ and $\epsilon$
between edges $2$ and $9$ become $\pi$. Similar statement holds for
edges $4$ and $8$. Switching to
the one-polygon representation of the same universe amounts to deleting
edge $3$ and the vertices on it (which no longer carry any deficit angles)
and considering $2$ and $9$ as well as $4$ and $8$ as single edges.}}
\end{center}
\end{figure}

In order to illustrate the relation between tessellations
with different numbers $F$ of polygons, 
let us consider an explicit example. 
Fig.\ref{tupoltri} represents a 
tessellation with $F=2$, corresponding to the piecewise flat
universe of Fig.\ref{tupol}. The figure is analogous to
Fig.\ref{ss} (before mapping it to $S$) with the
difference that 
the dual graph $\gamma$ (dotted lines) 
now has two distinct base points $P$ and $Q$ as indicated.
The solid lines represent edges of $\Gamma$. Any such edge
which reaches the boundary of the big quadrilateral must be glued
to the outgoing edge labeled by the same number. 
The two flat polygons
corresponding to this example are shown schematically in 
Fig.\ref{tupolfla}. Closed dual loops based at $P$ ($Q$) correspond to 
edges which appear twice in the polygon with centre $P$ ($Q$), 
and dual curves connecting $P$ to $Q$
correspond to edges appearing in both polygons. After the gluing the 
two polygons form a connected piecewise flat surface of genus 2.

Consider now the situation when one of the dual edges, say, edge
3, has zero length. The base points $P$ and $Q$ then fall on top of
each other, and the triangles $348$ 
and $392$ are degenerate, since the two edges $4$ and $8$ coincide, 
as do $2$ and $9$. The angles enclosed by these pairs of edges are zero,
and consequently the angles between the edges of $\Gamma$ 
(indicated in Fig.\ref{tupoltri}) on the piecewise flat surface
are $\pi$. Furthermore, since the boost parameter is zero, the matching
condition in the most general case is given by
\beq 
X_P=R(\phi)X_Q-a. 
\eeq       
This implies we can redefine $X_Q$ to be
exactly $X_P$ without changing the shape of the polygon
(since the transformation is a pure rotation). 
It also means that edge $3$ is redundant, and edges $4$ and $8$ (as well as $2$ 
and $9$) can be represented by just single edges. We have therefore
rederived the situation of Fig.\ref{onepol}.         

The intriguing conclusion from this analysis is
that the boost parameter of an edge bounding two
distinct polygons can {\it always} be made to vanish by an
appropriate gauge transformation. The general
matching condition of edge $3$ (omitting the translational part) 
reads
\beq 
X_P=\Lambda X_Q, \label{mcagain} 
\eeq
where $X_P$ and $X_Q$ are now distinct coordinate systems. 
After performing two independent Lorentz transformations on both 
frames ($X_P\to\tilde X_P=\Lambda_P X_P$ and
$X_Q\to\tilde X_Q=\Lambda_Q X_Q$) eq.(\ref{mcagain}) becomes
\beq 
X_P=\Lambda_P^{-1} \Lambda \Lambda_Q X_Q. 
\eeq
There are now many choices for $\Lambda_P$ and $\Lambda_Q$ 
which reduce the matching condition
to $X_P=X_Q$ (for example, $\Lambda_P=\Lambda$ and 
$\Lambda_Q=I$ will do),
which means that we can effectively get back to a one-polygon 
tessellation by performing a symmetry transformation. By iterating
this argument, one can reduce any multi-polygon to one equivalent
to a single polygon (cf. Appendix D), while leaving the smooth surface 
$S$ unchanged.

However, there is a caveat which prevents us from proving the
physical equivalence of any multi-polygon universe with a
one-polygon universe. Note that in performing the finite
Lorentz transformations of local Minkowski frames
to gauge-fix some of the $\eta$'s, we have so far completely
ignored how these transformations act on the shapes of the polygons. 
Although infinitesimal Lorentz transformations of this type can 
always be performed, for larger transformation one will in general
encounter the same transitions that can occur during time
evolution, and therefore be forced to switch to a new set of edge
variables.
For example, due to a concave vertex hitting an opposite edge, 
any one-polygon component may decompose into several pieces, 
thus defeating the aim of {\it reducing} the number of polygons.

In other words, we may not be able to ``lift'' the action of any 
given Lorentz transformation on the angle variables $(\eta_i)$ 
to one on the full set $(L_i,\eta_i)$ of phase space variables
for a given set {$e_i$} of edges. This potential obstruction
to reduction is not as bad as it may at first appear, 
since already at the
level of the $\eta$'s the reduction is highly non-unique. 
Unfortunately, the action of the symmetry group on the 
phase space is rather involved and to find elements in
$\times_F${\it PSU}(1,1) which avoid transitions which increase 
$F$ seems to
depend on the details of the geometry of the multi-polygon
universe. We have so far neither a general argument nor
an explicit algorithm for reducing any given universe 
to one equivalent to a one-polygon universe, but only a

\vspace{.3cm}
\noindent {\it Conjecture.} For any multi-polygon universe given in terms
of $6g+3(F-2)$ pairs of edge variables $(L_i,\eta_i)$, one
can always find a joint frame transformation in \newline
$\times_{j=1}^F${\it PSU(1,1)$_j$} which avoids transitions
and leads to a gauge-equivalent configuration with only 
$6g-3$ non-vanishing variable pairs, and physically equivalent
to a one-polygon universe.\footnote{A weaker version of the 
conjecture still sufficient would be to allow for an increase in $F$ 
at an intermediate stage of the construction.} 
\vspace{.3cm}

If the conjecture is true, the solution space $\cal P$
associated with one-polygon universes we constructed in Sec.3
coincides with the full physical reduced phase space of
pure (2+1) gravity for compact slices of genus $g$.

\section{Discussion}

By re-interpreting the boost parameters of 't Hooft's polygon model
as geodesic lengths on a hyperbolic surface $S$ of constant curvature $R=-1$
we have succeeded in finding an explicit parametrization of
a sector $\cal P$ of dimension $12g-12$ of the physical phase space 
of the model. The lengths are those of the geodesic loops and arcs of a 
triangulation of $S$, whose vertices are in one-to-one correspondence 
with the polygons of the physical Cauchy surface. At the level of
$S$, the action of the gauge transformations of the polygon model
is well understood, and corresponds to moving the vertices of the 
triangulation. 

If there is only a single vertex, the corresponding constant-time
slice of the (2+1) universe consists of one polygon. For this case,
we have solved the complex initial-value constraint of the 
model, and have given a complete parametrization of the
solution space $\cal P$ of initial data. This was done by 
expressing the boost parameters as functions of the Teichm\"uller
parameters of $S$, and then complementing them by suitable edge length
variables which solve the initial value constraint. 
We conjecture that $\cal P$ coincides with the 
full phase space of (2+1) vacuum gravity. Equivalently, we conjecture
that any multi-polygon tessellation is physically equivalent to
a one-polygon tessellation. 
We also showed that time evolution in $\cal P$ is either eternal expansion 
from a big bang or shrinking to a big crunch from the infinite past,
thus generalizing a known result from the genus-1 case.

A natural next step in our analysis will be the inclusion 
of point particles. In the case without particles treated in 
this paper, the generators $g_i$ are
hyperbolic elements of {\it PSU}(1,1) ($\vert$tr$\:g_i\vert>2$), which have
no fixed points in $D^2$. The generator corresponding to a particle on
the other hand will be elliptic
($\vert$tr$\:g_i\vert<2$). It has a fixed point in $D^2$ (since it is
conjugate to a rotation), and the corresponding
hyperbolic structure $D^2/G$ will have singular points as expected.  
We believe that the method introduced in this paper can be
generalized to this case, by substituting the smooth surface $S$
by one with a number of punctures, one for each particle.

Another issue one might like to investigate is 
the role played by the mapping class group G in the higher-genus
case. Its action is not difficult to describe:
a Dehn twist is obtained by conjugating the generators 
$g_i$ of $G$ by a group element $h \in G$.
What kind of further redundancies this may 
induce on the phase space and whether 
those can at all be ``factored out" are in general very difficult
issues. (For contradictory claims in the torus case, 
see \cite{mod1,mod2}.

This work -- motivated by having a computer code but being unable 
to produce a
set of initial data solving the constraints --   
is a step toward understanding the asymptotic behaviour
of 2+1 gravity near the singularity and other dynamical issues
about which very little is currently known for higher-genus universes. 
A numerical treatment is now feasible, but also analytical progress
may be within reach. A beautiful feature of the 't Hooft model is
the simplicity of its dynamics, since all the length variables $L$
evolve linearly. As we have seen, this picture is partially 
spoiled by the presence of ``transitions", points in the
evolution at which the number of phase space variables (before
imposing the constraints) jumps. If our conjecture is indeed true,
it would mean that such transitions can be completely avoided,
leading to a considerable simplification of both the classical and
potentially also the quantum theory.

\section*{Acknowledgements}
Z.K. thanks M. Carfora, G. 't Hooft, D. N\'ogr\'adi, B. Szendr\H oi
and D. Thurston for discussions. 
     
\appendix

\section{Boosts in the one-polygon tessellation}

In this appendix we prove that a one-polygon tessellation only admits
configurations where all boost parameters have the same
sign. 
Recall that 
$\pi < \sum_{i} \alpha_i<2\pi$ holds for mixed 
vertices $v$ of $\Gamma$, and $2\pi < \sum_{i }
\alpha_i<3\pi$ for the homogeneous ones. Since the graph
$\Gamma$ has $4g-2$ 
vertices, in the case
of two mixed vertices the sum of all angles in the polygon is given
by (cf. (\ref{constr1})) 
\beq 
(12g-8)\pi=\sum_{ i=1}^{12g-6} \alpha_i<4\pi+(4g-4)3\pi=(12g-8)\pi.
\label{inconsistent}
\eeq 
This is a contradiction, so we can exclude the appearance of 
more than one mixed vertex. Note that if the graph $\Gamma$ was
one-particle irreducible we would already be done, since there cannot 
be only a single mixed vertex in such a graph. 

Suppose then that one vertex $v_0$ is mixed. For the sum of 
the angles of all hyperbolic triangles we have 
\beq 
\sum_{i=1}^{12g-6}\tilde\alpha_i=2\pi, 
\label{sumtilde}
\eeq
since all of them contribute at the base point of the
triangulation $\gamma$, and this point does not carry
a non-trivial deficit angle.
Next, recalling that at the mixed vertex 
$\alpha_3=\tilde{\alpha}_3+\pi$ and $\alpha_1=\tilde{\alpha}_1,\;
\alpha_2=\tilde{\alpha}_2$, 
the polygon closure constraint is given by
\beq 
\begin{array}{l} {\displaystyle \sum_{i=1}^{12g-6} 
\alpha_i=\sum_{i\;{\rm not\, at}\, v_0}
(\pi-\tilde{\alpha}_i)+\tilde{\alpha}_1+\tilde{\alpha}_2+
\tilde{\alpha_3}+\pi=}
\\
{\displaystyle
=(4g-3)3\pi+\pi+\sum_{i\; {\rm at}\, v_0    }\tilde{\alpha_i}-
\sum_{i\; {\rm not\, at}\, v_0} 
\tilde{\alpha_i}=(12g-8)\pi}.
\end{array}
\label{manyalpha}
\eeq
The last equality is the requirement of the constraint. 
Relation (\ref{manyalpha}) implies that
\beq 
\sum_{i\; {\rm at}\, v_0}
\tilde{\alpha}_i-\sum_{i\; {\rm not\, at}\, v_0}
\tilde{\alpha}_i=0, 
\eeq 
which in conjunction with eq.(\ref{sumtilde}) leads to
\beq
\sum_{i\; {\rm at}\, v_0 }\tilde{\alpha_i}=\pi
\eeq
for the triplet of angles associated with $v_0$.
This means that the corresponding hyperbolic 
triangle is degenerate since its area vanishes,
$\pi-\sum_i \tilde{\alpha_i}=0$. 
As a consequence, either two of its sides coincide and
the third one has length zero, or the union of its two short sides with 
angle $\pi$ between them coincides with the long side. 
The first case is again excluded 
since $\eta_i=0$ implies two vertices having no deficit angle
which again leads to an
inconsistency in (\ref{inconsistent}). (Although the contribution
of the mixed vertices on the right-hand side is maximized to $4\pi$,
the homogeneous vertices still contribute with angle sums
strictly smaller than $3\pi$.)
The second case corresponds to the situation where
$\alpha_i=0$ for some $i$, which is not allowed.

\section{Boost parameters from Teichm\"uller space}

In this appendix we will explain how to obtain an independent
set of initial values for the boost parameters $\eta$. 
For simplicity and illustrative purposes, we will discuss an 
example of genus 2. Since we will not make use of the symmetry
structure of this particular case, the generalization to 
higher genus is immediate.

We fix the triangulation by choosing the graph $\Gamma$ on
Fig.\ref{onepol}, leading to the triangulation of
Fig.\ref{onepolcur}.\footnote{Recall
that the procedure described at the beginning of Sec.\ref{secc} 
gives the mirror image of Fig.\ref{onepolcur}.} We will use
the convention of multiplying loops from left to right. The
numbers indicate the outgoing ends of the loops, and $i=1,\dots,4$
label the generators $b_i$ of the fundamental
group satisfying
$b_1 b_2 b_1^{-1} b_2^{-1} b_3 b_4 b_3^{-1} b_4^{-1}=1$.
The homotopy classes of the remaining closed curves can be
obtained by composing the fundamental generators and their
inverses, leading to\footnote{Our and Okai's \cite{okai}
convention for multiplying elements (curves) of the fundamental
group is from right to left, so the product $v_1v_2$ in 
$\pi_1(S)$ in our case is mapped to $g_2g_1$ in the corresponding
Lie group.}
\beq 
\begin{array}{ccl}
5 & \to & b_2 b_1 b_2^{-1} b_1^{-1} \\
6 & \to & b_1 b_2^{-1} \\
7 & \to & b_2 b_1 b_2^{-1} \\
8 & \to & b_3 b_4^{-1} \\
9 & \to & b_4 b_3 b_4^{-1}. 
\end{array}
\eeq
Now we use the faithful representation of $\pi_1(S)$ in {\it PSU}$(1,1)$ 
given explicitly in \cite{okai} in terms of the so-called
Fenchel-Nielsen coordinates. They are a set of length and angle
variables $(l_k,\tau_k)$, $k=1,\dots,3g-3$, which parametrize the 
Teichm\"uller space ${\cal T}_g$ globally. One can pick an 
arbitrary element
$(l,\tau) \in {\mathbb R}_{+}^{3g-3}\times{\mathbb
  R}^{3g-3}\cong{\cal T}_g$, plug it into the formulae for the
generators $g_i \in$ {\it PSU}(1,1) and compute the combinations for the
group elements corresponding to the remaining curves in the 
triangulation (in our specific example, the curves labeled 
5, 6, 7, 8 and 9). Now,
knowing the group generators and the combinatorial information 
(the order of the edges going around the polygon), we
can identify the boost parameters and compute the angles as well.
How the boosts are supplemented by a set of
length variables $L_i$ has been described in Sec.3.6 above.
Altogether this amounts to an explicit algorithm 
for constructing a set of initial data for a 
(2+1)-dimensional universe from any element of 
$ {\mathbb R}_+^{6g-6}\times {\cal T}_g $.

\section{The complex constraint}

The proof that for a one-polygon tessellation 
there is always a Lorentz frame\footnote{Equivalently,
a suitable base point in $S$, or a suitable $h \in$ {\it PSU}(1,1) to
conjugate all the generators with.} in which the
complex constraint (\ref{comcon2}) admits a solution rests on
the following facts.
\begin{enumerate}
\item The complex vector $z_i$ defined below eq.(\ref{comcon2}) 
points to the angle bisector of the geodesic loop corresponding to $g_i$.
\item The velocity at $P$ of the unique arc connecting the base point 
$P\in S$ to the unique smooth closed geodesic corresponding to 
$g_i$ points in the same direction.  
\item One can find a base point where these velocities are not
contained in a half plane.
\item In practice one proceeds by finding an element 
$h \in${\it PSU}(1,1) which
corresponds to the desired change of base point. The original vectors 
$z_i$ can be read off from $g_i \in G$. The new coefficients $z_i'$ 
will be determined from the conjugated generators $h^{-1}g_ih$, and 
they will not lie in a half plane by the above arguments.   
\end{enumerate}
Take the universal cover where the origin $0\in D^2$ is mapped to the
base point $P\in S$. The in- and out-going ends of the loop $g_i$ 
on $S$ can be associated with
the geodesic arcs connecting $0$ with $g_i0$ and $0$ with
$g^{-1}_i0$ (the group action on points $z\in D^2$ 
was defined in eq.(\ref{su11})). Since the
geodesics through $0\in D^2$ are Euclidean straight lines, 
and since
\beq 
g_i0=\tanh \eta_i\exp i \phi_i, \quad g_i^{-1}0=\tanh \eta_i \exp
i (\pi-\phi'_i),  
\label{bisect}
\eeq
it is clear that the angle bisector of $g_i0$ and $g^{-1}_i0$ 
points in the same direction as $z_i$.

\begin{figure}
\begin{center}
\includegraphics{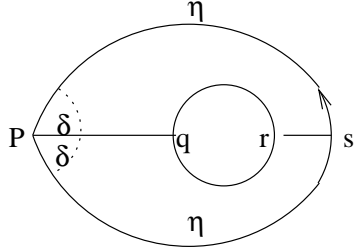}
\caption{\label{trirectangle} {On the
surface $S$, the loop connecting the basepoint $P$ to itself via point
$s$ is the loop in the triangulation corresponding to $g_i$. The
circle in the middle is the unique smooth geodesic which lies
in the same homotopy class as the loop. The unique arcs connecting 
$P$ to the circle and the circle to the loop such that the angles at 
$q$, $r$ and $s$ are right angles 
create two isometric quadrilaterals.}}   
\end{center}
\end{figure}

\begin{figure}
\begin{center}
\includegraphics{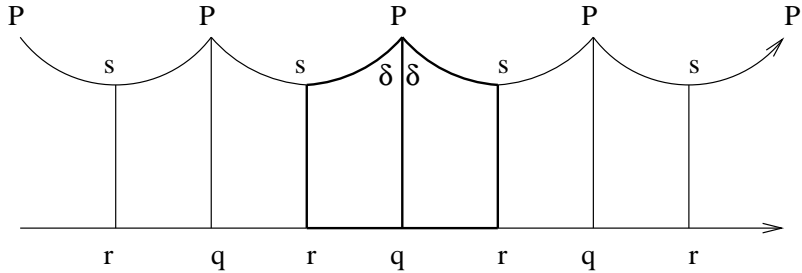}
\caption{\label{trirect2} {Lift to $D^2$ of a (periodically extended) 
geodesic loop at $P$ (top) and of the associated homotopic smooth 
geodesic (bottom). The geodesic arc $rs$ is the unique perpendicular
connecting the geodesic loop $PP$ to the smooth closed geodesic $qq$. 
The unique geodesic arc connecting $P$ to $qq$ which is perpendicular
at $q$ is the angle bisector of the loop and its inverse at $P$. 
The thick lines indicate one of the (infinitely many) pairs
of isometric quadrilaterals $Pqrs$ (each with three right
angles) which are mapped to 
Fig.\ref{trirectangle} under the universal covering map $f$.}}   
\end{center}
\end{figure}

The next step is to establish the validity of Fig.\ref{trirectangle},
namely, that the two quadrilaterals $Pqrs$ are isometric.
The figure shows the smooth geodesic (inner circle) freely homotopic 
to the geodesic loop $PP$ and the unique smooth
geodesic arcs $Pq$ and $rs$ connecting two, and perpendicular to 
them in the points $q$, $r$ and $s$. All we need to show is that
the arc $Pq$ is the angle bisector of the velocities of the in-
and outgoing ends of the loop (denoted by $\eta$ in the figure).
We refer to \cite{buser} for a detailed proof.
The properties of the various geodesics in $D^2$ are illustrated
by Fig.\ref{trirect2}. The smooth geodesic of Fig.\ref{trirectangle}
is mapped to the smooth straight line on the bottom, and the
geodesic loop to the periodic non-smooth curve at the top.
The curves at $q$, $r$ and $s$ meet at right angles. 


The situation on $D^2$ is as follows. 
We are given the set of $6g-3$ elements $\{g_i\}$ of the
fundamental group $G$. 
Each $g_i$ has its so-called {\it axis}, that is, the geodesic 
which is left invariant by $g_i$. On the disk $D^2$, an axis
has the form of a circle segment whose ends are perpendicular 
to the disk boundary. (This fact is not reproduced in 
Fig.\ref{trirect2}, where the axis is represented by the straight 
line at the bottom.) Under the universal cover, an axis is mapped 
(infinitely many times) to the smooth
geodesic on $S$ corresponding to $g_i$.
The unique geodesic arc on $D^2$ 
from the origin which is perpendicular to one of these axes 
is mapped to the arc on $S$ connecting the base point to
the smooth geodesic in question. 

\begin{figure}
\begin{center}
\includegraphics[width=9cm,height=5.75cm]{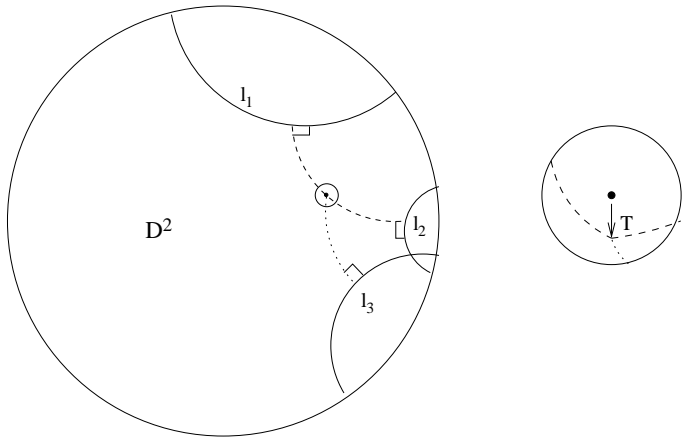}
\caption{\label{disk} {The circle segment $l_i$ perpendicular
to the disk boundary is the so-called axis of $g_i$.
In the figure, $g_1$ and $g_2$ correspond to two non-intersecting 
smooth geodesics. A good new base point $T$ is either given
by the mid point of the unique arc orthogonal to
both $l_1$ and $l_2$ (dashed line), or by moving slightly away from the 
mid point and from the arc, as explained in the text.
In the figure, after moving $T$ to a location below the arc
(as indicated in the magnified picture  
of the neighbourhood of the mid point on the right),
the three initial velocities of the unique geodesics connecting
$T$ with $l_1$, $l_2$ and $l_3$ (and perpendicular to them)
will span the entire two-plane.
}}   
\end{center}
\end{figure}

Suppose now that their initial velocities at the origin 
are contained in a half plane (otherwise we are already done). 
We can then find a point $T \in D^2$, such that the initial velocities 
of the unique arcs -- emanating from $T$ and perpendicular to 
the axes -- do not lie in a half plane. The procedure for
finding such a location is straightforward, and is
illustrated in Fig.\ref{disk}. Pick two loops $g_1$ and $g_2$ 
based at $P$
whose associated smooth geodesic loops do not 
intersect\footnote{Such loops always exist, see eg. \cite{mosher}.}.
Consider the mid point $M$ of the unique geodesic arc perpendicular 
to both of the associated axes, $l_1$ and $l_2$. If we were to
choose $M$ as the new base point, the two oppositely oriented
tangent vectors to the arc at $M$ would define the directions of
the new complex vectors $z_1'$ and $z_2'$. 
There are then two possibilities. (i) The remaining axes
$l_i$, $i\geq 3$, do not lie just to one side of the
arc, so that their tangent vectors at $M$ already span the
entire two-plane. In this case we are done and $T\equiv M$ is 
a good new base point. (ii) The remaining axes lie to one side
of the arc only, so that their tangent vectors, together with
$z_1'$ and $z_2'$ span only a half plane. For example, in 
Fig.\ref{disk} they all lie like $l_3$, that is, below the
arc. In that case it will be sufficient to move the point
to a new point $T$
slightly away from the arc, in the direction of the
remaining axes. The new tangent vectors $z_1''$ and $z_2''$
will then enclose an angle slightly smaller than $\pi$, and
together with $z_3''$, say, span the entire two-plane.
In either case, a suitable new base point on $S$ is the
image of $T$ under the universal cover. For calculational
purposes, it is convenient to change the universal cover to another one
whose origin is mapped to the new base point. The effect of this is
to conjugate $g_i \to h^{-1}g_ih$. The
triangulation in $D^2$ determined by the set $\{g_i\}$ with base point
$T$ is isometric to that determined by $\{h^{-1}g_ih\}$ with base point
$0$, with
\beq 
T \to h^{-1}T=0,\quad g_iT \to h^{-1}g_ih0=h^{-1}g_iT. 
\eeq
We have thus completed the proof that one can always find a Lorentz
frame in which the complex constraint admits a solution.

\section{Eliminating polygons by gauge-fixing}

In this appendix we will show how to gauge-transform 
a given $(\tilde{\gamma},F)$ (a geodesic triangulation 
$\tilde\gamma$ of some $F$-polygon) to a configuration
$(\tilde{\gamma}',F)$ which is equivalent to 
a configuration $(\tilde{\gamma}'',F-1)$ with one
polygon fewer.\footnote{For the purposes of this appendix, we
will mean by $\tilde\gamma$ a geodesic triangulation together
with a definite length assignments $\eta_i$ to its edges, and
by $\gamma$ the underlying topological triangulation.}
The induced map $(\tilde{\gamma}',F)\mapsto
(\tilde{\gamma}'',F-1)$ 
amounts to deleting
three edges and one base point from $\tilde{\gamma}'$ 
but does not change $\tilde{\gamma}'$ as a point set. 

A gauge transformation of $(\tilde{\gamma},F)$ is an action of
$\times_F${\it PSU}(1,1), where each of the $F$ 
copies of {\it PSU}(1,1)$\equiv$
{\it SO}(2,1) acts independently as follows. If $\tilde e_{ij}$ denotes
an oriented edge connecting base points $\tilde P_i$ and $\tilde P_j$ on $S$, 
we will call its associated group element $g_{ij}
\equiv g_{ji}^{-1}$. To give an example, for the edge $\tilde e_{12}$ connecting
$\tilde P_1$ and $\tilde P_2$, we have 
$\bar{P_2}=g_{21}\bar{P_1}$ for the inverse 
images in $D^2$. A generic gauge transformation is given by an 
$F$-tuple $(h_1,h_2,\dots,h_F)\in \times_F${\it PSU}(1,1), acting
by group multiplication at the end points of edges according to
\beq
g_{ij}\mapsto h_i g_{ij}h_j^{-1},\quad i,j\in \{1,2,\dots,F\}.
\label{groupmult}
\eeq
There will usually be several edges linking a base point to itself
(implying $i=j$), which can be taken care of by introducing an
extra label for the edges and group elements, $e_{ii}^{(k)}$ and 
$g_{ii}^{(k)}$.
At the level of the frames $X_j$, $1<j<F$, and assuming for the
moment no obstructions,
this gauge transformation corresponds to a 
simultaneous rotation of the frames, $X_j\rightarrow \tilde X_j=
\Lambda_j X_j$ via the canonical isomorphism $h_i\sim\Lambda_i$
of subsection 3.5.
If two neighbouring frames were related by a Lorentz transformation
$\Lambda_{21}$ before the gauge transformation, $X_2=\Lambda_{21} X_1$,
the matching condition afterwards will be
$\tilde X_2=\tilde\Lambda_{21} \tilde X_1$, with
$\tilde\Lambda_{21}=\Lambda_2\Lambda_{21}\Lambda_1^{-1}$,
cf. (\ref{groupmult}).

Let us adopt the notation $\bar P$ for points in $D^2$ and
$\tilde P$ for their images in $S$ under the universal cover,
and suppose that $h_i\bar P_i\in D^2$ is mapped to
$\tilde P_i'$, $i=1,2$. Then the 
boost parameter $2\eta_{21}'$
read off from the group element $h_2^{-1}g_{21}h_1$ is the length of 
the geodesic arc
connecting $\tilde P_1'$ to $\tilde P_2'$, which is freely homotopic 
to the original arc
connecting $\tilde P_1$ to $\tilde P_2$ with length $\eta_{21}$. 
We conclude that also in generic multi-polygon universes
gauge transformations amount to moving the base points without changing 
the topology of the graph $\gamma$.     

Suppose now that we perform a gauge transformation on a single frame
only, say, $X_1$. The effect on the geometry of the graph 
$\tilde\gamma$ will
be a motion of the base point $\tilde P_1$ and a modification of 
the edges starting or ending at $\tilde P_1$. The magnitude of
the change will be chosen as $g_{21}$, corresponding to the
geodesic arc $\tilde e_{21}\in \tilde\gamma$ connecting 
$\tilde P_1$ to $\tilde P_1'=\tilde P_2$. 
Its effect can be written as follows:
\beq \begin{array}{llll} \label{gfix}
\tilde e_{11}^{(i)}:\tilde P_1 \to \tilde P_1& \mapsto & 
\tilde e_{11}^{(i)'}:\tilde P_1' \to \tilde P_1', &   
g_{11}^{(i)} \mapsto g_{21}g_{11}^{(i)}  g_{21}^{-1} \\
\tilde e_{k1}^{(i)}:\tilde P_1 \to \tilde P_k, & \mapsto & 
\tilde e_{k1}^{(i)'}:\tilde P_1' \to \tilde P_k, & 
g_{k1}^{(i)} \mapsto g_{k1}^{(i)} g_{21}^{-1} \\ 
\tilde e_{1k}^{(i)}:\tilde P_k \to \tilde P_1, & \mapsto & 
\tilde e_{1k}^{(i)'}:\tilde P_k \to \tilde P_1', & 
g_{1k}^{(i)} \mapsto g_{21}  g_{1k}^{(i)} \\ 
\tilde e_{kl}^{(i)}: \tilde P_l \to \tilde P_k, & \mapsto & 
\tilde e_{kl}^{(i)'}:\tilde P_l \to \tilde P_k, & 
g_{kl}^{(i)} \mapsto g_{kl}^{(i)}, \\
\end{array} 
\eeq
assuming $k,l\neq 1$. 
Writing $\tilde P_1'$ in (\ref{gfix}) is meant to emphasize that despite
$\tilde P_1'=\tilde P_2$ one has to keep track of whether an end point
of a curve in $\tilde{\gamma}$ corresponds to the base point labeled
by $1$ or by $2$. 
Consider now one of the two triangles in $\tilde{\gamma}$ which 
share the geodesic arc $\tilde e_{21}$ (we have dropped the counting
label $i$ for simplicity). 
It consists of the arcs $\tilde e_{21}:\tilde P_1 \to \tilde P_2$,
$\tilde e_{1k}:\tilde P_k \to
\tilde P_1$ and $\tilde e_{2k}:\tilde P_k\to \tilde P_2$, 
and we have $g_{2k}=g_{21}g_{1k}$ for the
corresponding group elements. The action of the above transformation on
these arcs and group elements reads
\beq 
\begin{array}{llll}
\tilde e_{21}  :\tilde P_1 \to \tilde P_2 & \mapsto & 
\tilde e_{21}':\tilde P_1' \to \tilde P_2, & 
g_{21}\mapsto g_{21}g_{21}^{-1}=1 \\ 
\tilde e_{1k}  :\tilde P_k \to \tilde P_1 & \mapsto & 
\tilde e_{1k}':\tilde P_k \to \tilde P_1'=\tilde P_2, &  
g_{1k}\mapsto g_{21}g_{1k}=g_{2k} \\ 
\tilde e_{2k}  :\tilde P_k \to \tilde P_2 & \mapsto & 
\tilde e_{2k}':\tilde P_k \to \tilde P_2, &   g_{2k}\mapsto g_{2k}.
\end{array} 
\eeq
In other words, arc $\tilde e_{21}$ has shrunk to length zero (the trivial curve),
arc $\tilde e_{1k}$ has been transformed to coincide with $\tilde e_{2k}$, 
and arc $\tilde e_{2k}$ has
been left untouched. The new geodesic triangle with sides 
$\tilde e_{21}'$, $\tilde e_{1k}'$ and $\tilde e_{2k}'$
is degenerate. The same is true for the other triangle that shared the
edge $\tilde e_{21}$.
In order to obtain the reduced graph $(\tilde\gamma'',F-1)$ from
$(\tilde\gamma',F)$, we delete the
redundant base point $\tilde P_1'$ and arc $\tilde e_{21}'$,
as well as one arc of the pair $(\tilde e_{1k}',\tilde e_{2k}')$,
and one arc from the corresponding pair of the neighbouring triangle.
Note that $\tilde{\gamma}''=\tilde{\gamma}'$
as point sets, but that $\tilde{\gamma}'$ has one trivial and two 
double edges.

\end{document}